# Double Quantum Well Triple Barrier Structures: Analytical and Numerical Results

[1]R. V. Ramos and [2]S. D. G. Martinz
rubens.viana@pq.cnpq.br samueldgm@fisica.ufc.br

[1]*Federal Institute of Education, Science and Technology of Ceara, Fortaleza-Ce, Brazil.*
[2]*Lab. of Quantum Information Technology, Department of Teleinformatic Engineering – Federal University of Ceara - DETI/UFC, C.P. 6007 – Campus do Pici - 60455-970 Fortaleza-Ce, Brazil.*

***Abstract* — In this work we show an analytical result for the scattering in a particular type of double quantum well triple barrier structure and numerical results, via the Numerov method, for bound states of a double quantum well triple barrier inside of a infinite quantum well. For the last, we consider both, constant and position dependent mass.**

## I. Introduction

In the design of new nanoelectronic devices, to find the solutions of the Schrödinger equation for arbitrary potential profiles is a crucial step. In general, one is looking for the transmission (or reflection) coefficient of a scattering nanostructure, or bound solutions for particles trapped in wells (wires or dots). Due to the hardness of finding analytical solutions for arbitrary potentials, numerical methods for solving the Schrodinger equation have been largely employed [1-4]. In what concerns the scattering, the simplest case, found in any quantum mechanics text book, is the single-barrier with height $V_0$ and width $a$, as shown in Fig. 1.

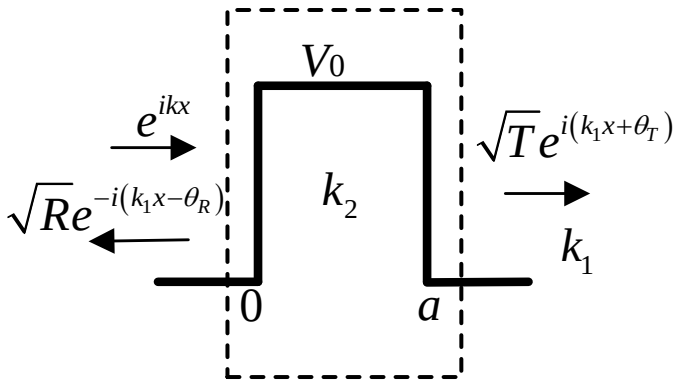

Fig. 1 – Single barrier with height $V_0$ and width $a$.

It can be easily shown that the transmission and reflection coefficients of the barrier in Fig. 1 are given by

$$t = \frac{2k_1k_2e^{-ik_1a}}{2k_1k_2\cos(k_2a) - i(k_1^2+k_2^2)\sin(k_2a)} \qquad (1)$$

$$r = \frac{(k_1^2-k_2^2)\sin(k_2a)}{(k_1^2+k_2^2)\sin(k_2a) + 2ik_1k_2\cos(k_2a)} \qquad (2)$$

$$k_1^2 = 2mE/\hbar^2,\quad k_2^2 = 2m(E-V_0)/\hbar^2. \qquad (3)$$

In (3) $m$ is the particle's mass while $E$ is its energy. After some manipulations of (1) and (2) one readily finds the transmissivity and reflectivity

$$T = |t|^2 = \frac{4E(E-V_0)}{4E(E-V_0) + V_0^2\sin^2\left[\frac{\sqrt{2m(E-V_0)}}{\hbar}a\right]} \qquad (4)$$

$$R = |r|^2 = \frac{V_0^2\sin^2\left[\left(\sqrt{2m(E-V_0)}/\hbar\right)a\right]}{4E(E-V_0) + V_0^2\sin^2\left[\left(\sqrt{2m(E-V_0)}/\hbar\right)a\right]}. \qquad (5)$$

as well the transmission and reflection angles

$$\theta_T = -k_1a - \tan^{-1}\left[-\frac{(k_1^2-k_2'^2)}{2k_1k_2'}\tanh(k_2'a)\right] \qquad (6)$$

$$\theta_R = \frac{\pi}{2} + \tan^{-1}\left[\frac{(k_1^2-k_2'^2)}{2k_1k_2'}\tanh(k_2'a)\right]. \qquad (7)$$

Climbing the complexity scale, the next case that can be found in text books of nanoelectronic devices is the single-quantum-well double barrier (SQW-DB) structure shown in Fig. 2 [5,6].

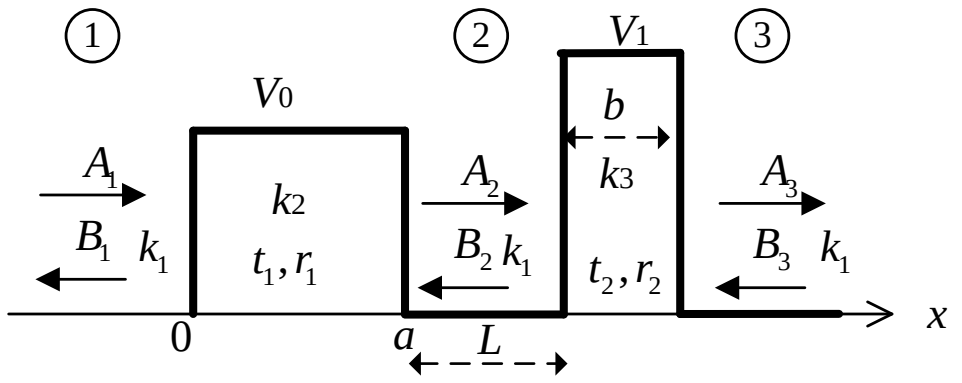

Fig. 2 – Single-quantum-well double-barrier structure.

After some calculations one can find that the transmission coefficient of the nanostructure in Fig. 2 is given by

$$t = t_1t_2e^{ik_1L}\Big/\left(1 - r_1r_2e^{i2k_1L}\right) \qquad (8)$$

where $t_1$ and $r_1$ are given by (1) and (2), respectively, while $t_2$ and $r_2$ are given by

$$t_2 = 2k_1k_3e^{-ik_1b} \Big/ \left[2k_1k_3\cos\left(k_3b\right) - i\left(k_1^2 + k_3^2\right)\sin\left(k_3b\right)\right] \quad (9)$$

$$r_2 = \frac{\left(k_1^2 - k_3^2\right)\sin\left(k_3a\right)}{\left(k_1^2 + k_3^2\right)\sin\left(k_3b\right) + 2ik_1k_3\cos\left(k_3b\right)} \quad (10)$$

$$k_3^2 = 2m\left(E - V_1\right)/\hbar^2. \quad (11)$$

In the case of two identical barriers (8) is reduced to

$$t = t_1^2 e^{ik_1L} \Big/ \left(1 - r_1^2 e^{i2k_1L}\right) \quad (12)$$

and, hence, its transmissivity is [7]

$$T \equiv |t|^2 = \left[1 + 4\left(R_1/T_1^2\right)\sin^2\left(k_1L + \theta_1^R\right)\right]^{-1}. \quad (13)$$

In (13), $R_1$ and $T_1$ are given by (4) and (5), respectively, while $\theta_1^R$ is given by (7).

Rising one step more in the complexity scale, one finds the double quantum well triple barrier (DQW-TB) structure. Such structure has found application, for example, in resonant tunneling devices [8,9]. In this direction, the present work brings a formula for the transmission coefficient of a particular type of DQW-TB structure and numerical results, by using the Numerov Method, for the bound states (and associated energies) of the DQW-TB inside of an infinite quantum well.

This work is outlined as follows: In Section 2, we provide a formula for the transmission coefficient of a particular type of DQW-TB nanostructure. In Section 3 we show numerical results of a DQW-TB inside of an infinite quantum well. Section 4 considers the case of a DQW-TB inside of an infinite quantum well with mass dependent position. At last, the conclusions are drawn in Section 5.

## II. Double Quantum Well Triple Barrier Nanostructure

Double quantum well triple barrier nanostructure has being studied since the 1990's [10]. Since then some works about DQW-TB have been done by using numerical methods, mainly the transfer matrix method [10-12]. Here, for the first time in the best of our knowledge, we present an analytical formula for the transmissivity of the (rectangular) DQW-TB with equal lateral barriers, as shown in Fig. 3.

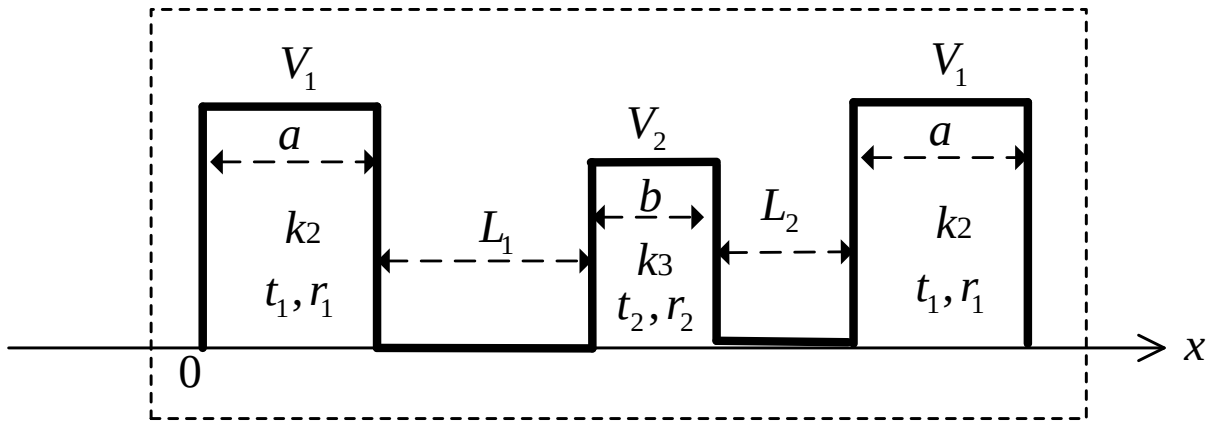

Fig. 3 – Double quantum well triple barrier structure.

The transmission coefficient of the nanostructure in Fig. 3 is given by

$$t = \frac{t_1^2 t_2}{2r_1r_2\cos\left(k_1\left(L_2 - L_1\right)\right)e^{ik_1\left(L_1+L_2\right)} - 1 + \left(t_2/t_2'\right)r_1^2e^{i2k_1\left(L_1+L_2\right)}} \quad (14)$$

where $t_1$ and $r_1$ are given, respectively, by (1) and (2), $t_2$ and $r_2$ are given, respectively, by (9) and (10) and

$$t_2' = 2k_1k_3e^{-ik_1b} \Big/ \left[2k_1k_3\cos\left(k_3b\right) + i\left(k_1^2 + k_3^2\right)\sin\left(k_3b\right)\right] \quad (15)$$

$$k_1^2 = 2mE/\hbar^2, \;\; k_2^2 = 2m\left(E - V_1\right)/\hbar^2, \;\; k_3^2 = 2m\left(E - V_2\right)/\hbar^2 \quad (16)$$

In Fig. 4 it is shown the transmissivity of the DQW-TB for a particle's mass of $0.067m_e$, calculated by using (14)-(16) (I) and numerically by using the transfer matrix method (II) as well the transmissivity of the SQW-DB (III) formed by the same DQW-TB without the central barrier. The following parameters' values were used for curves (I) and (II): $L_1 = L_2 = 2.5$nm, $a = 2.5$nm, $b = 1.5$nm, $V_1 = 0.4655$eV and $V_2 = 0.3258$eV. For curve (III) it was used $L_1 = L_2 = 3.25$nm, $a = 2.5$ nm, $b = 0$nm, $V_1 = 0.4655$eV and $V_2 = 0$eV. Observing Fig. 4 one can readily note the correctness of our analytical result given in (14)-(16). Furthermore, for the set of parameters' values used, the central barrier causes the presence of a narrow high transmission peak ($T = 0.9646$ at 0.1529eV) that does not exist in the transmissivity of the SQW-DB nanostructure. Hence, at 0.1529eV the DQW-TB is a bad reflector while the SQW-DB is a good reflector. The opposite occurs at 0.5396eV, when $T = 0.9974$ for SQW-DB while $T = 0.1313$ for DQW-TB. One can note that an electron with energy 0.5396eV can be used to distinguish a SQW-DB from a DQW-TB in an interaction free measurement [13].

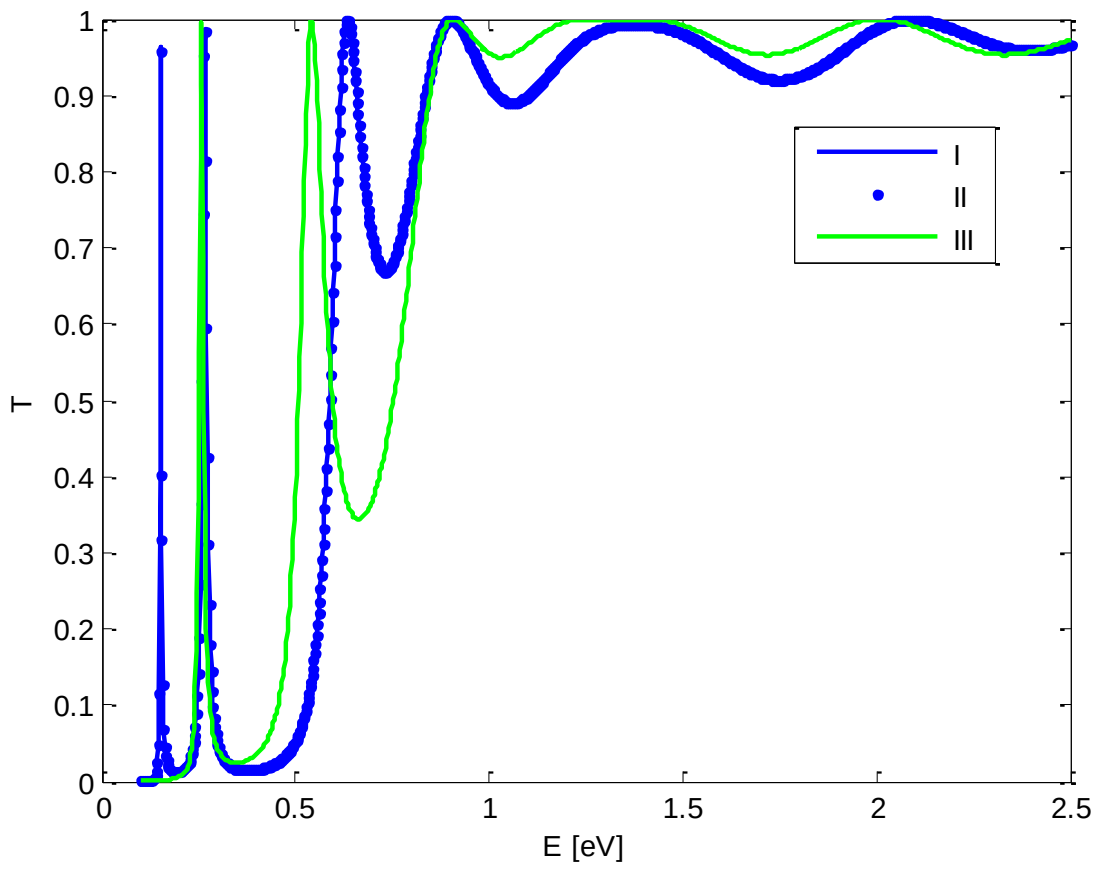

Fig. 4 – Transmissivity versus energy for a DQW-TB: I – Analytical result given by (14)-(16); II – Numerical result obtained by using the transfer matrix method. III - Transmissivity of the SQW-DB structure formed by the DQW-TB without the central barrier.

In Fig. 5 one can see the transmissivity of the DQW-TB structure for I) $V_1$= 1eV and $V_2$ = 0.5eV; II) $V_1$= 1eV and $V_2$ = 2eV. In both cases it was used $L_1$ = $L_2$ = 2.5nm, $a$ = 2.5nm and $b$ =1.5nm.

Finally, one may note in (14) that once $L_1$+$L_2$ and |$L_1$-$L_2$| have been fixed, the transmissivity does not change. For example, the transmissivity for $L_1$ = 1nm and $L_2$ = 4nm is the same when $L_1$ = 4nm and $L_2$ = 1nm.

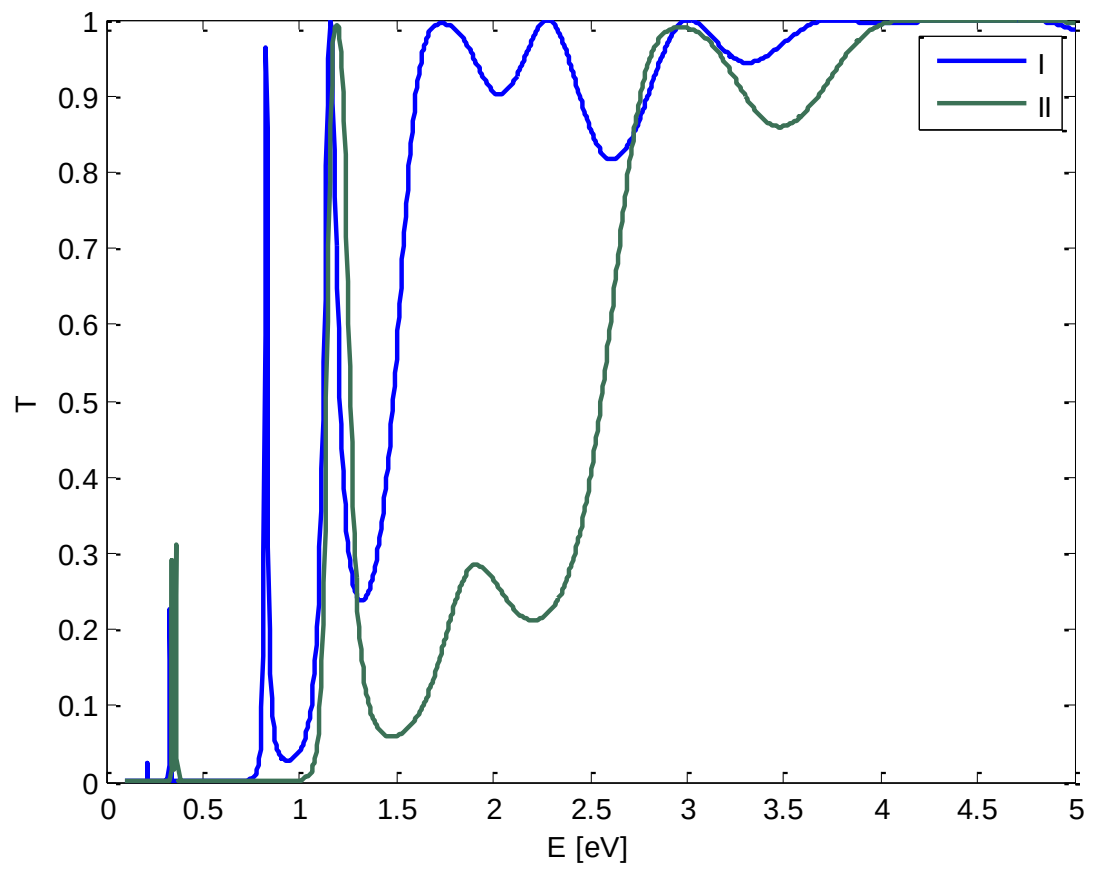

Fig. 5 – Transmissivity versus energy for a DQW-TB. I) $V_1$= 1eV and $V_2$= 0.5eV; II) $V_1$= 1eV and $V_2$= 2eV.

## III. Double-Quantum-Well Triple-Barrier Inside of an Infinite Quantum Well

Hereafter, we consider the problem of a DQW-TB inside of an infinite quantum well. In order to find the bound states and the associated energies, one has to solve the time independent 1D Schrödinger equation

$$d^2\psi(x)/dx^2 = -\left(2m/\hbar^2\right)\left[E - V(x)\right]\psi(x). \tag{17}$$

Its discretization using the Numerov method is [14]

$$\left[-\frac{\hbar^2}{2m}B^{-1}A + V\right]\psi = E\psi \tag{18}$$

$$\psi = \left[\ldots \quad \psi_{n-1} \quad \psi_n \quad \psi_{n+1} \quad \ldots\right]^T \tag{19}$$

$$A = \left(I_{-1} - 2I_0 + I_1\right)/\left(\Delta x\right)^2 \tag{20}$$

$$B = \left(I_{-1} + 10I_0 + I_1\right)/12 \tag{21}$$

$$V = \begin{bmatrix} V(x_1) & 0 & 0 \\ 0 & \ddots & 0 \\ 0 & 0 & V(x_n) \end{bmatrix} \tag{22}$$

In (20)-(21) $I_r$, $r$ = 0, 1, and -1, is a matrix of 1s along the $r$th diagonal and zeros elsewhere.

Hereafter, in all simulations with rectangular barriers, the DQW-TB having $a$ = $b$ = $L_1$ = $L_2$ = 3nm is placed inside of an infinite quantum well with width equal to 21 nm. The particle's mass is once more 0.067$m_e$. Furthermore, the eigenfunctions are multiplied and/or displaced by a constant factor in order to permit their visualization inside of the potential function profile. Some eigenfunctions solutions of (18)-(22) for the DQW-TB inside of an infinite quantum well, can be seen in Figs. 6 ($V_1$ = $V_2$ = 1eV), 7 ($V_1$ = $V_2$ = 3eV) and 8 ($V_1$ = $V_2$ = 5eV).

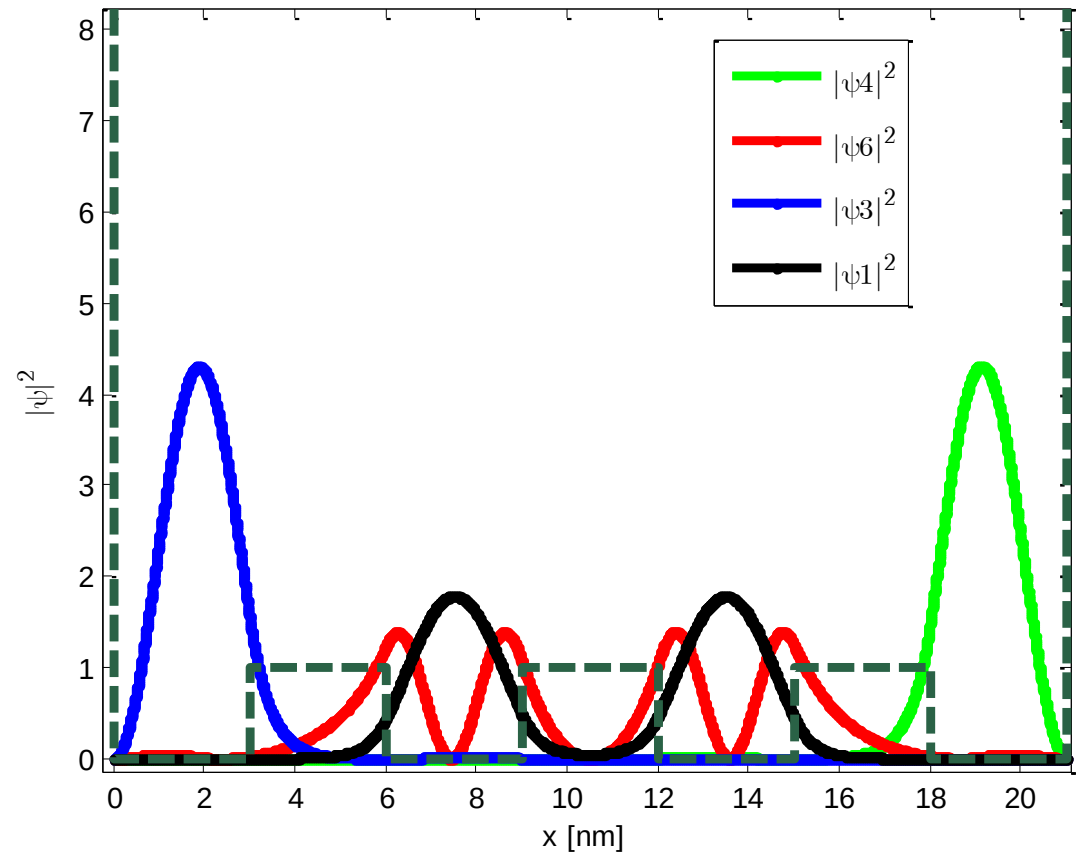

Fig. 6 – The squared modulus of the eigenfunctions $\psi_1$, $\psi_3$, $\psi_4$ and $\psi_6$. The parameters values are $V_1$ = $V_2$ = 1eV. The energies of the modes are $E_1$ = 0.2449eV, $E_3$ = 0.3522eV, $E_4$ = 0.3537eV, $E_6$ = 0.9025eV.

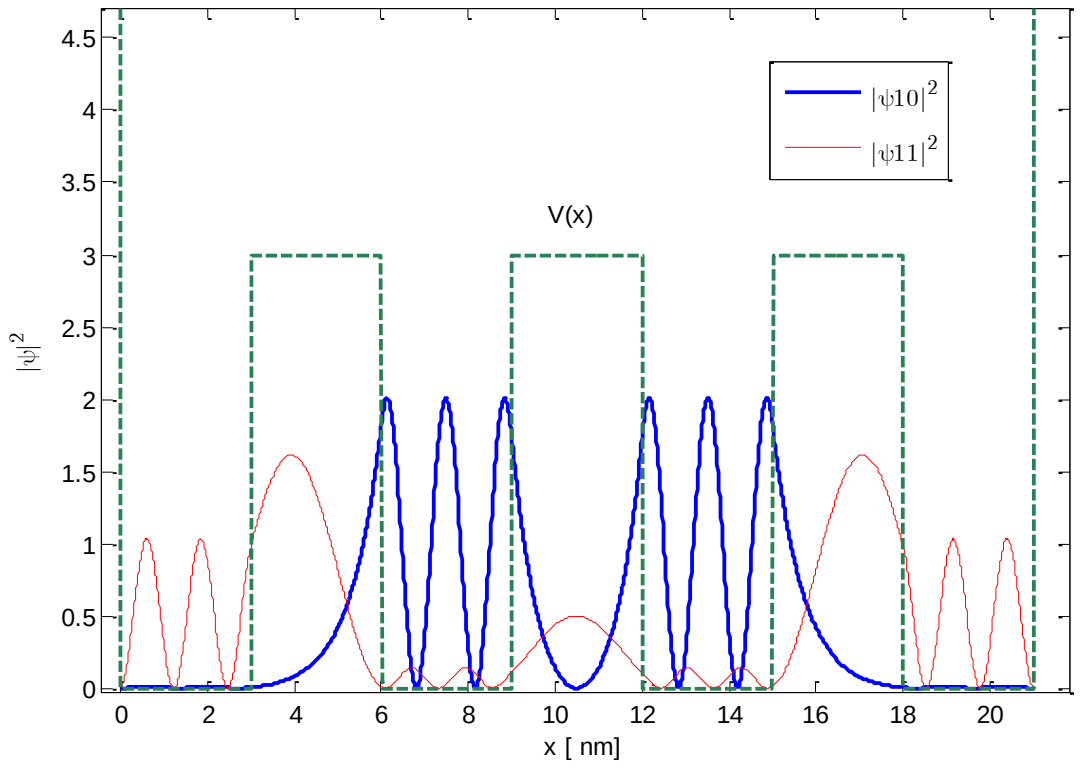

Fig. 7 – The squared modulus of the eigenfunctions $\psi_{10}$ and $\psi_{11}$. The parameters values are $V_1 = V_2 = 3$eV. The energies of the modes are $E_{10} = 2.7435$eV and $E_{11} = 3.2672$eV.

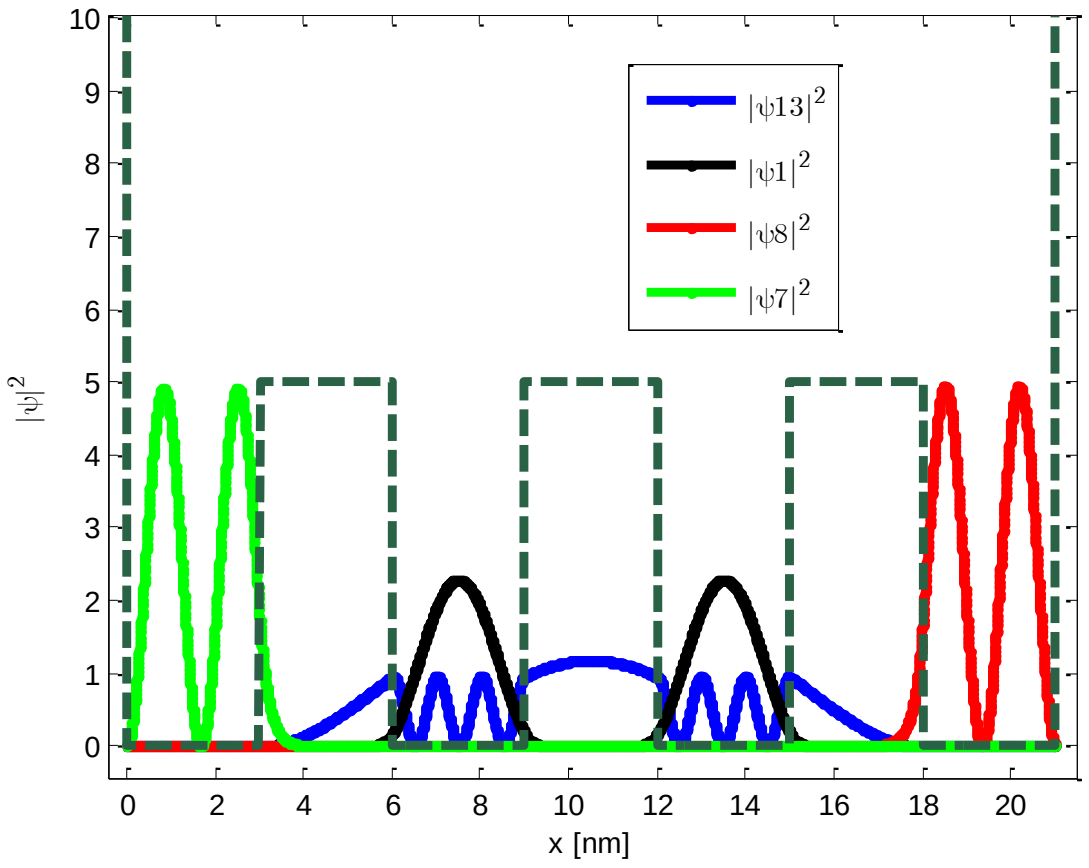

Fig. 8 – The squared modulus of the eigenfunctions $\psi_1$, $\psi_6$, $\psi_7$ and $\psi_{13}$. The parameters values are $V_1 = V_2 = 5$eV. The energies of the modes are $E_1 = 0.3774$eV, $E_7 = 1.7842$eV, $E_8 = 1.7930$eV and $E_{13} = 5.0469$ eV.

As can be noted, the particle can be partially confined between the two lateral barriers of the DQW-TB ($\psi_1$ and $\psi_6$ in Fig. 6, $\psi_{10}$ in Fig. 7, $\psi_1$ and $\psi_{13}$ in Fig. 8), between a barrier and a wall of the infinite quantum well ($\psi_3$ and $\psi_4$ in Fig. 6, $\psi_6$ and $\psi_7$ in Fig. 7) or not confined by the DQW-TB, what occurs when the energy is larger than the height of the barriers, for example $\psi_{11}$ in Fig. 7. Interestingly, there not exist a mode in which the particle keeps confined only between the central barrier and a lateral barrier (an eigenfunction with non-negligible amplitude only in the region 6nm $\le x \le$ 9nm or in the region 12nm $\le x \le$ 15nm). This means that even in the lowest energy mode the particle can tunnel through the central barrier with high probability. For example, in Fig. 8 one has $E_1 = 0.3774$eV for $\psi_1$ while the barrier height is 5 eV.

Another interesting situation occurs when the barriers location inside the quantum well are the most probable places to find the particle. This is the case for the mode $n$ = 15, which has energy (5.4929 eV) larger than the height of the barriers (5 eV), as shown in Fig. 9. In this case, the barriers are partially confining the particle, instead of the wells.

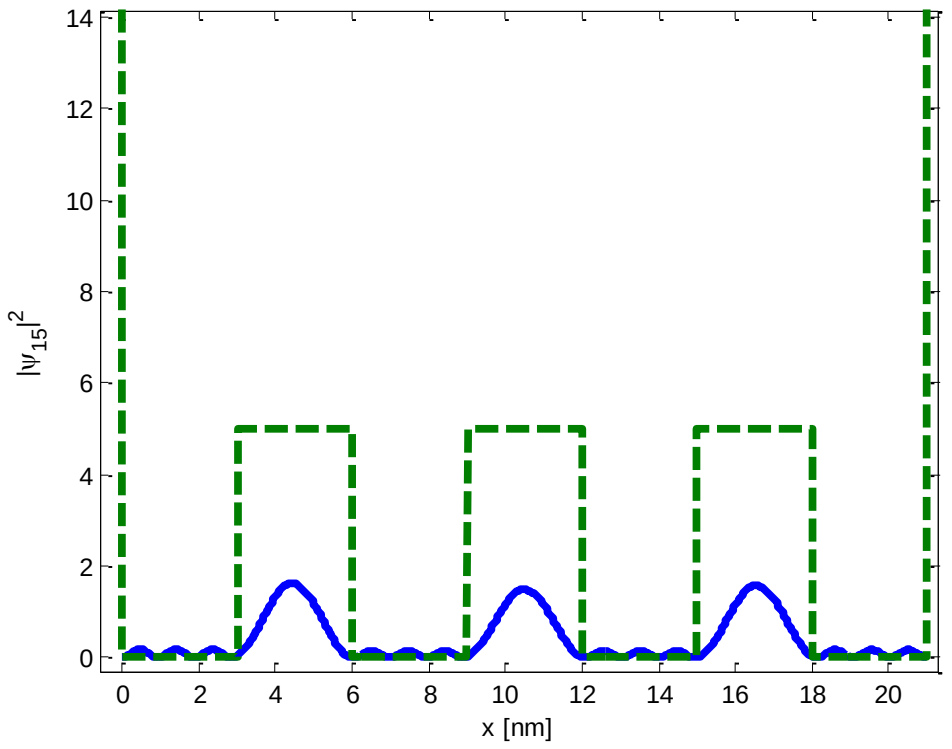

Fig. 9 – The squared modulus of the eigenfunction $\psi_{15}$. The parameters values are $V_1 = V_2 = 5$eV. The energy of the mode is $E_{15} = 5.4929$eV.

Furthermore, as expected, the lower the height of the barriers, the lower is the number of partially confined modes and vice-versa. The energies of the first ten modes for five different barriers' height are shown in Fig. 10.

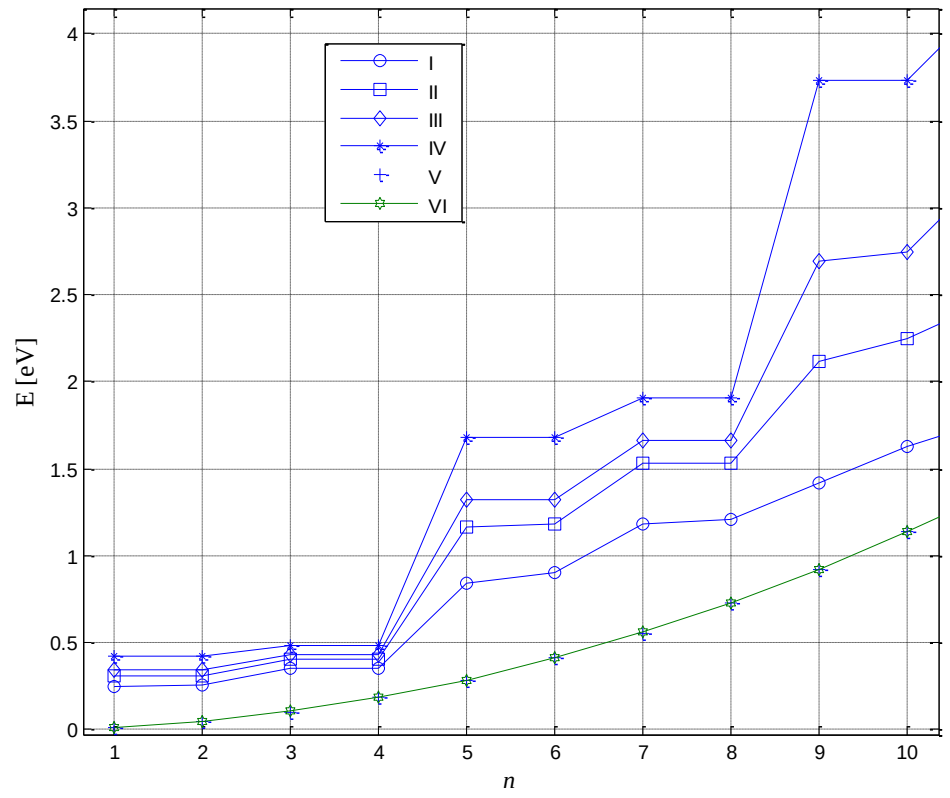

Fig. 10 – Energies of the first ten modes: I) $V_1 = V_2 = 1$eV; II) $V_1 = V_2 = 2$eV; III) $V_1 = V_2 = 3$eV; IV) $V_1 = V_2 = 10$eV; V) $V_1 = V_2 = 0$eV; VI) $E = n^2\hbar^2\pi^2/2mL^2$.

As can be seen in Fig. 10, the presence of the barriers inside the infinite quantum well breaks the parabolic behavior of the energy found in an infinite quantum well (curves V and VI). Moreover, the larger the barriers' height the larger is the energy of the modes. In order to conclude the study of DQW-TB with equal barriers, we consider a DQW-TB with (super)Gaussian barriers inside of an infinite quantum well having width equal to 20nm. The potential function inside the quantum well is

$$V(x) = \sum_{i=1}^{3} V_i \exp\left[-3(x-5i)^{\alpha}\right] \qquad (23)$$

In Fig. 11 one can see the squared modulus of the eigenfunctions $\psi_1$, $\psi_3$ and $\psi_7$ for $\alpha = 2$, and in Fig. 12 the energies of the first 15 modes for the following $\alpha$'s values: [2,6,10,14,18,22,26,30,34,38].

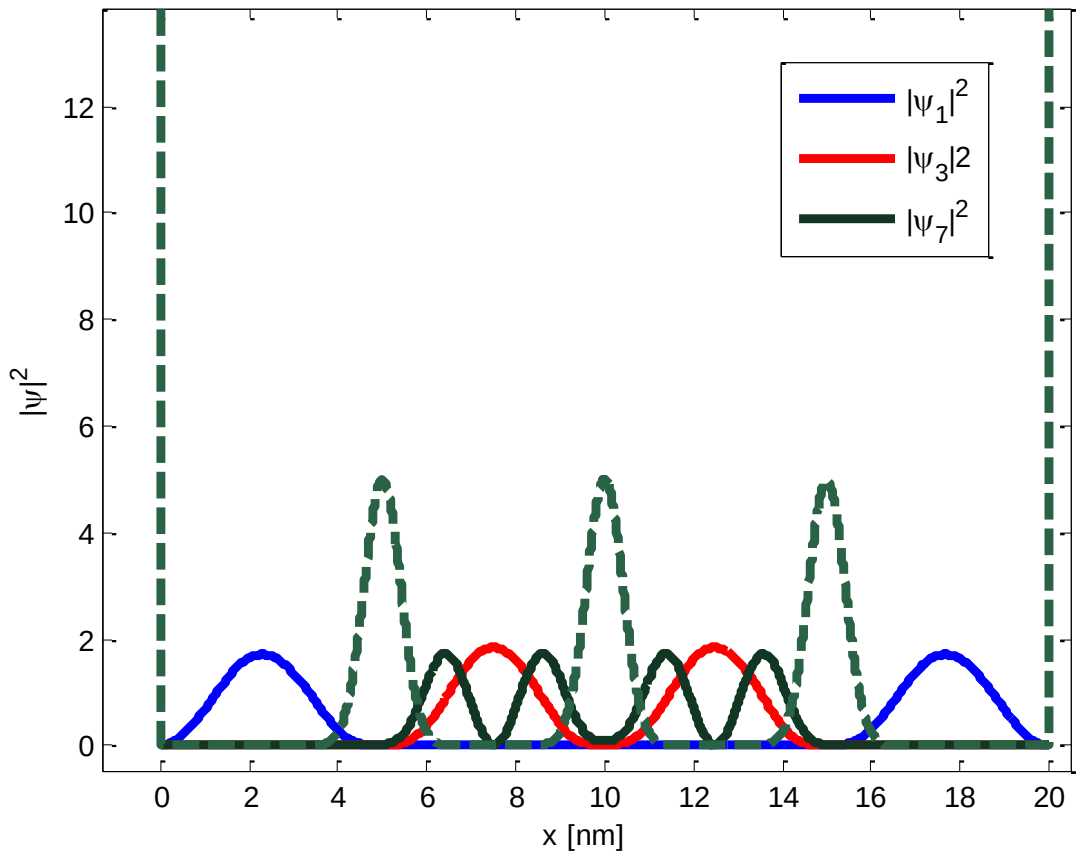

Fig. 11 – The squared modulus of the eigenfunctions $\psi_1$, $\psi_3$ and $\psi_7$. The parameters values are $V_1 = V_2 = 5$eV and $\alpha = 2$ in (23). The energies of the modes are $E_1 = 0.2333$eV, $E_3 = 0.2740$eV and $E_7 = 1.0540$eV.

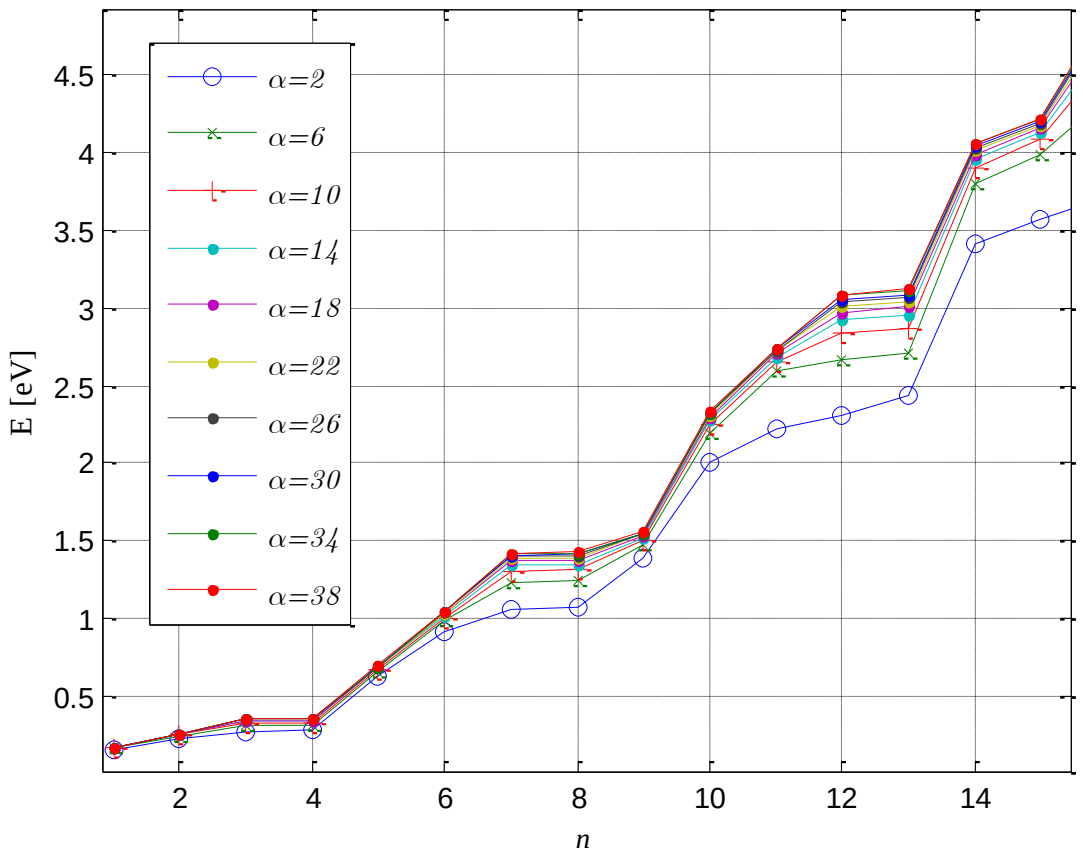

Fig. 12 – Energies of the first fifteen modes of the DQW-TB with supergaussian barriers inside of an infinite quantum well. The parameters values are $V_1 = V_2 = 5$eV and $\alpha$ = [2,6,10,14,18,22,26,30,34,38] in (23).

Now, considering the case of a DQW-TB with asymmetric barriers, Fig. 13 shows some eigenfunctions for a case in which the central barrier's is lower than the lateral barriers.

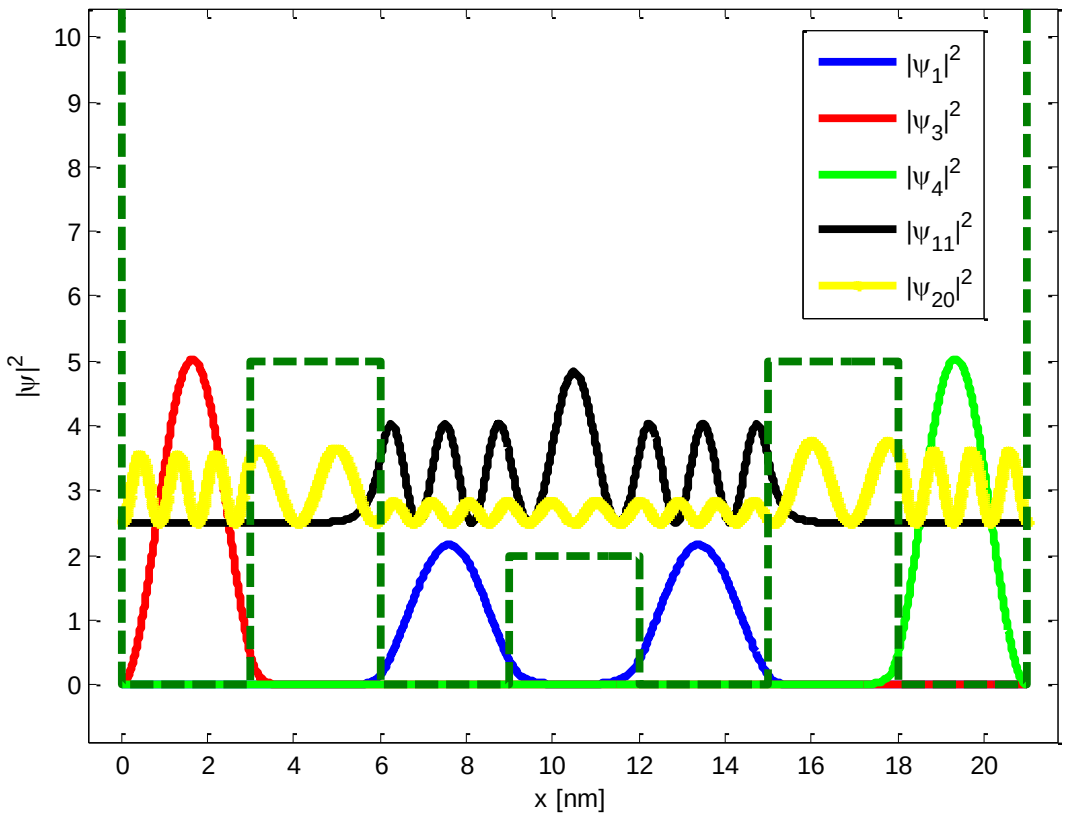

Fig. 13 – The squared modulus of the eigenfunctions $\psi_1$, $\psi_2$,$\psi_3$, $\psi_{11}$ and $\psi_{20}$. The parameters values are $V_1 = 5$eV, $V_2 = 2$eV. The energies of the modes are $E_1 = 0.3402$eV, $E_3 = 0.4516$eV, $E_{11} = 3.204$eV and $E_{20} = 6.5684$eV.

In Fig. 13, the eigenfunction $\psi_1$ represents a particle partially confined inside the quantum well formed by both lateral barriers, with low probability of being found in the region of the central barrier. The eigenfunction $\psi_3$ ($\psi_4$) represents a particle partially confined in the left (right) well formed by the wall of the infinite quantum well and the lateral barrier of the DQW-TB nanostructure. The eigenfunction $\psi_{11}$ represents a particle partially confined inside the quantum well formed by both lateral barriers but its energy is larger than the central barrier's height and, hence, the particle can be found in the central barrier region. Finally, the eigenfunction $\psi_{20}$ represents a particle confined inside the infinite quantum well and its energy is larger than any barrier's height.

Similarly to Fig. 9, in Fig. 14 it is shown the situation in which the particle is partially confined in the (lateral) barriers region.

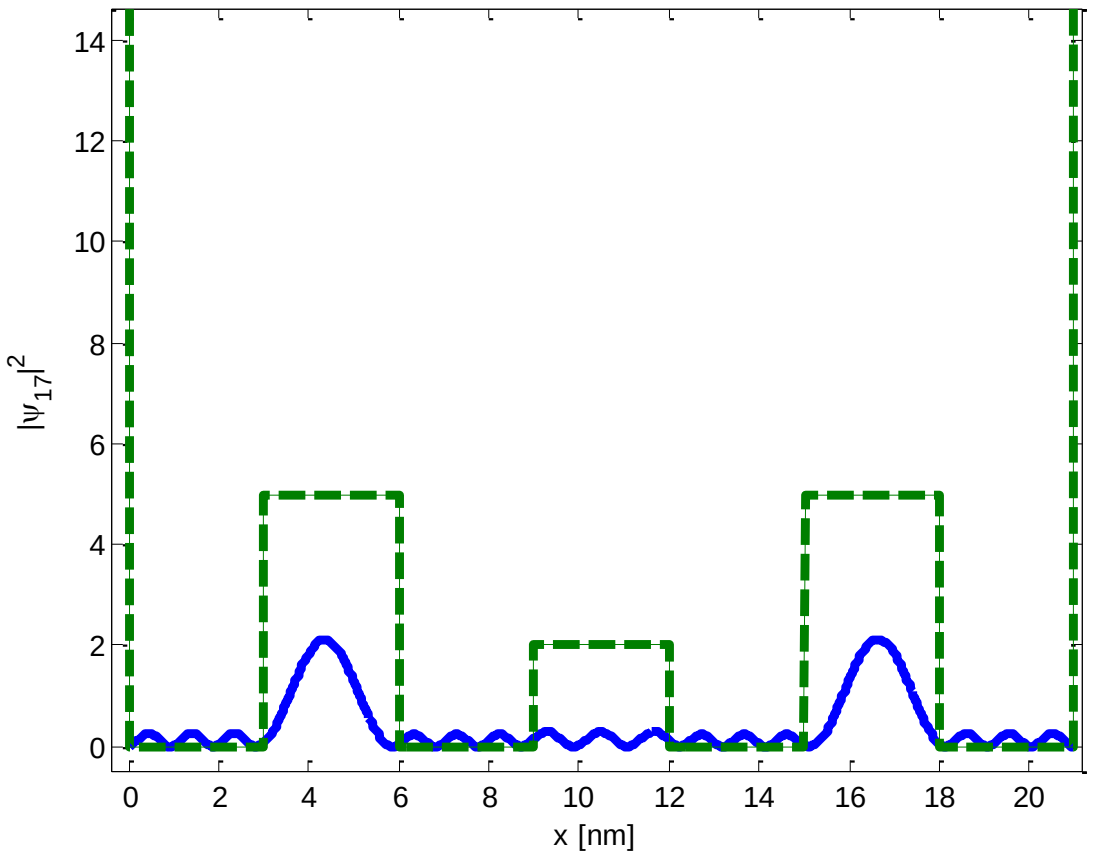

Fig. 14 – The squared modulus of the eigenfunctions $\psi_{17}$. The parameters values are $V_1 = 5$eV, $V_2 = 2$eV. The energy of the mode is $E_{17} = 5.5417$eV.

Figure 15 shows the energies for the first fifteen modes for three different situations of asymmetrical DQW-TB with central barrier lower than lateral barriers.

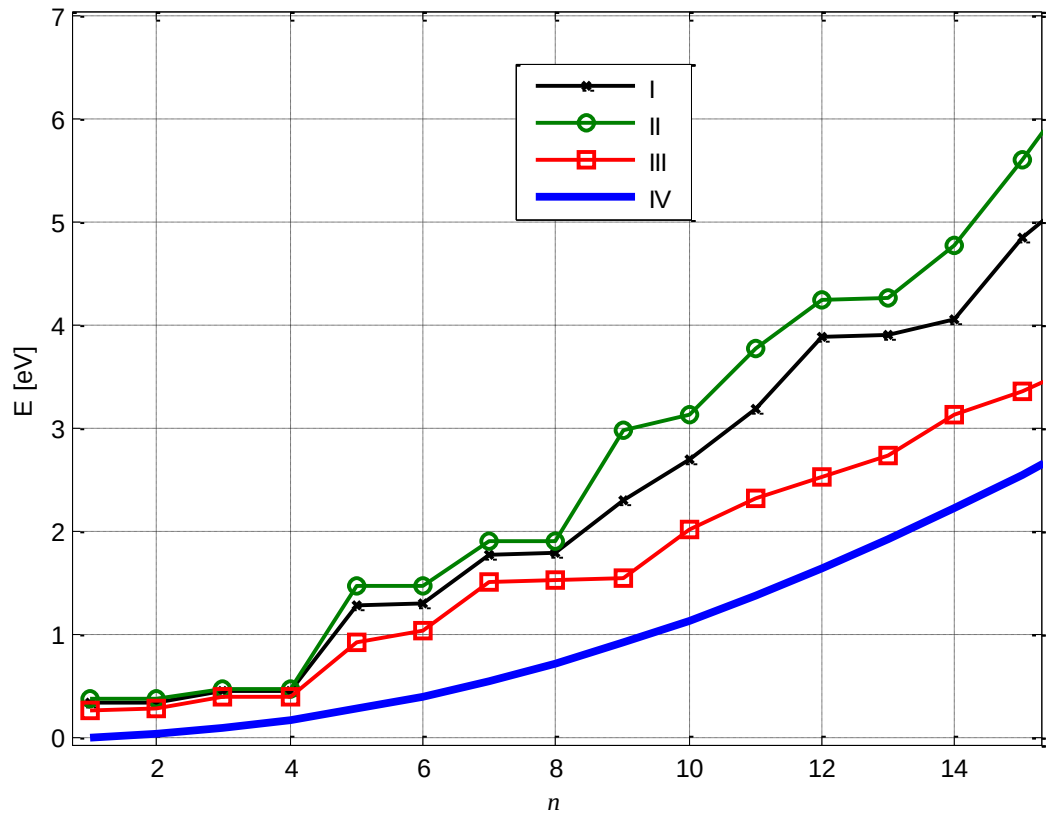

Fig. 15 – Energies of the first fifteen modes. The parameters values are: I) $V_1 = 5$eV and $V_2 = 2$eV; II) $V_1 = 10$eV and $V_2 = 3$eV; III) $V_1$=2eV and

$V_2$=1eV; IV) $E=n^2\hbar^2\pi^2/2mL^2$ ($m$ = 0.067$m_e$, $L$ = 21nm).

Finally, we consider the case in which the central barrier's height is larger than the lateral barriers' height. In Fig. 16 one can see some eigenfunctions of the DQW-TB inside of the infinite quantum well.

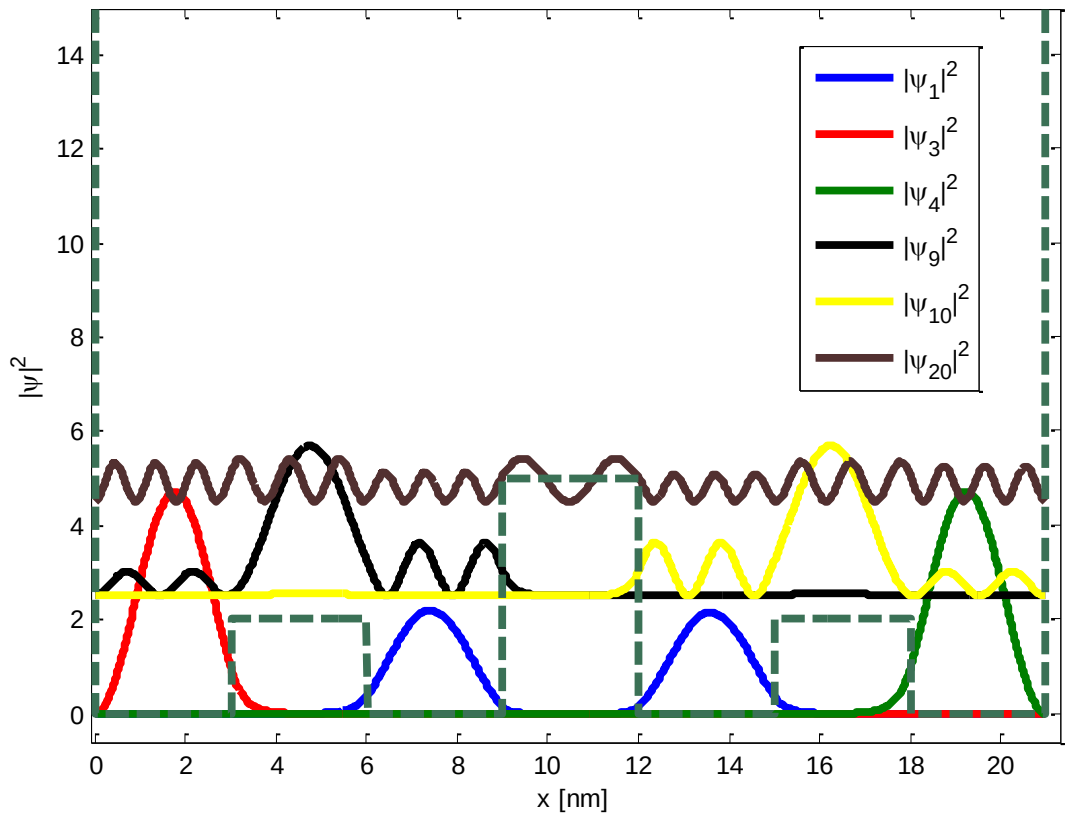

Fig. 16 – The squared modulus of the eigenfunctions $\psi_1$, $\psi_3$, $\psi_4$, $\psi_9$, $\psi_{10}$ and $\psi_{20}$. The parameters values are $V_1$ = 2eV, $V_2$ = 5eV. The energies of the modes are $E_1$ = 0.3399eV, $E_3$ = 0.4015eV, $E_4$ = 0.4034eV, $E_9$ = 2.3678eV, $E_{10}$ = 2.3690eV and $E_{20}$ = 6.1776eV.

The newness in Fig. 16 when compared to Fig. 13 is the presence of partially confined modes, $\psi_9$ and $\psi_{10}$, in the quantum well with a single barrier inside formed by the wall of the infinite quantum well, a lateral barrier (playing the role of central barrier) and the central barrier of the DQW-TB nanostructure. The case where the wavefunction gets more concentrated in the region of a barrier does also exist as shown in Fig. 17. In Fig. 18 one can see the energies of the first fifteen modes for three different situations of asymmetrical DQW-TB with central barrier higher than the lateral barriers.

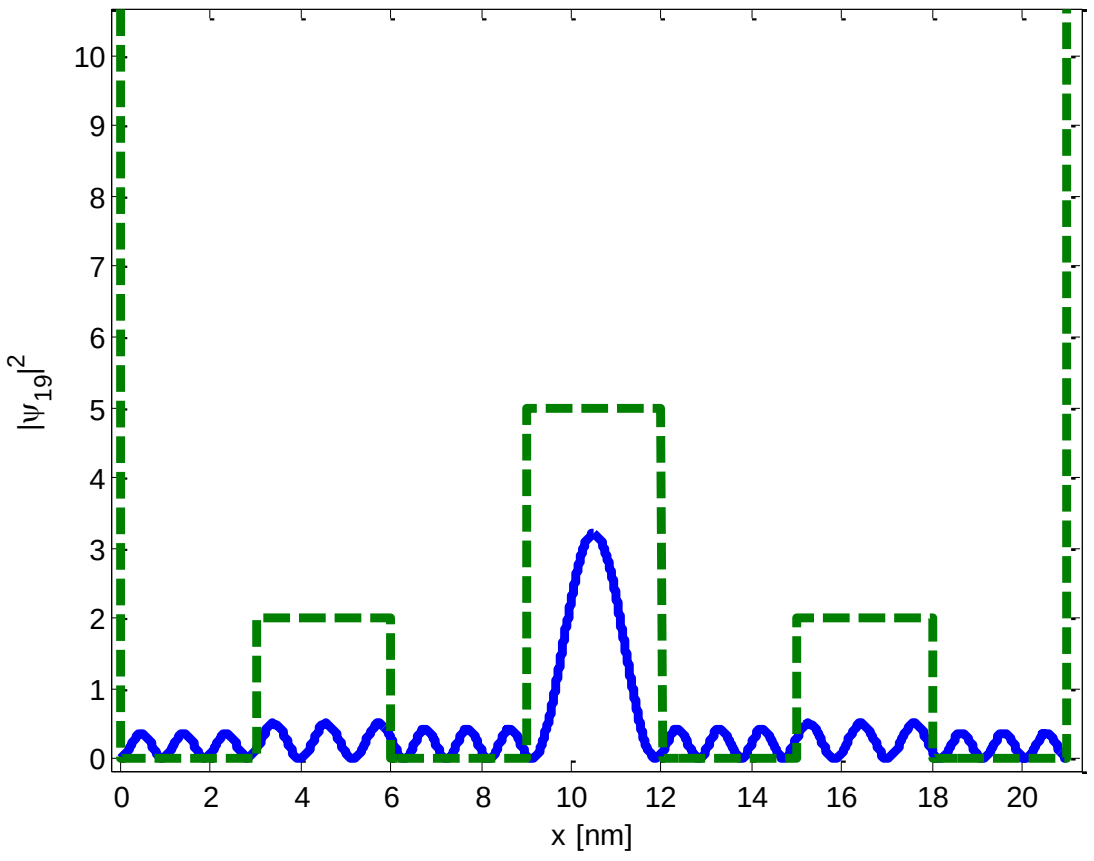

Fig. 17 – The squared modulus of the eigenfunctions $\psi_{19}$. The parameters values are $V_1$ = 2eV, $V_2$ = 5eV. The energy of the mode is $E_{19}$ = 5.6595eV.

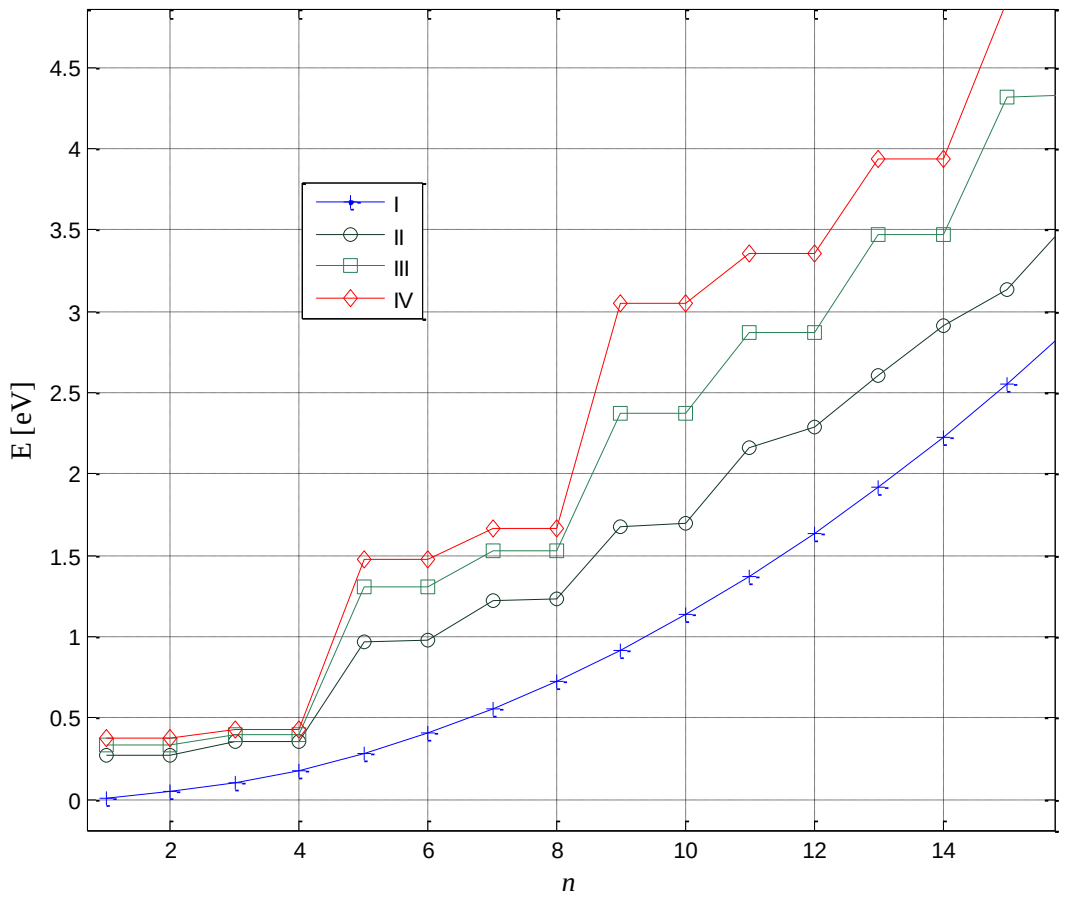

Fig. 18 – Energies of the first fifteen modes. I) $E=n^2\hbar^2\pi^2/2mL^2$ ($m$ = 0.067$m_e$, $L$ = 21nm); II) $V_1$ = 1eV, $V_2$ = 2eV; III) $V_1$ = 2eV, $V_2$ = 5eV; IV) $V_1$ = 3eV, $V_2$ = 10eV.

## IV. Double Quantum Well Triple Barrier Inside of an Infinite Quantum Well with Position Dependent Mass

When the particle's mass varies with position, the mass and momentum operators do not commute anymore. Hence, the kinetic energy operator must be modified. The kinetic operator proposed by Von Ross [15] is

$$T(x)=\frac{1}{4}\left(m^{\alpha}pm^{\beta}pm^{\gamma}+m^{\gamma}pm^{\beta}pm^{\alpha}\right) \tag{24}$$

$$\alpha+\beta+\gamma=-1. \tag{25}$$

Substituting the operator $p$ = $-i\hbar d/dx$ in (24), the Schrödinger equation is rewritten as [16]

$$\frac{d^2\psi}{dx^2}-\frac{m'}{m}\frac{d\psi}{dx}+\left[\frac{1}{2}\left(t\frac{m''}{m}-s\frac{m'^2}{m^2}\right)+\frac{2m}{\hbar^2}\left(E-V(x)\right)\right]\psi=0 \tag{26}$$

where

$$t=\alpha+\gamma,\quad s=\alpha(\gamma+2)-\gamma(\alpha+2) \tag{27}$$

$$m'=dm/dx,\ m''=d^2m/dx^2. \tag{28}$$

Now, making the substitution

$$\psi(x)=\sqrt{m(x)}\phi(x) \tag{29}$$

in (26), the Schrödinger equation can be rewritten as

$$\frac{\hbar^2}{2m_e}\frac{d^2\phi}{dx^2} = m_r(x)\left[E - V_{eff}(x)\right]\phi = 0 \tag{30}$$

$$m_r(x) = m(x)/m_e \tag{31}$$

$$V_{eff}(x) = V(x) - \left[\hbar^2(1+t)\frac{m''}{4m^2} - \hbar^2\left(\frac{3}{4}+\frac{s}{2}\right)\frac{m'^2}{2m^3}\right]. \tag{32}$$

Its discretization using the Numerov method is [17]

$$\left[-\frac{\hbar^2}{2m_e}M^{-1}A + V\right]\psi = E\psi \tag{33}$$

$$M = \left(M_{-1} + 10M_0 + M_1\right)/12 \tag{34}$$

$$M_{-1} = \begin{bmatrix} 0 & & & \\ m_r(x_1) & 0 & & \\ & \ddots & \ddots & \\ & & m_r(x_{n-1}) & 0 \end{bmatrix} \tag{35}$$

$$M_1 = \begin{bmatrix} 0 & m_r(x_2) & & \\ & 0 & \ddots & \\ & & \ddots & m_r(x_n) \\ & & & 0 \end{bmatrix}. \tag{36}$$

In (33) the matrix $A$ is given by (20), $V$ is a diagonal matrix whose entries are the values of $V_{eff}(x)$ and $M_0$ is a diagonal matrix whose entries are the values of $m_r(x)$. One may note that for a mass dependent position without very fast changes or discontinuities, the term inside of square brackets in the right side of (32) has a very low value even for very high values of $r$ and $s$ and, hence, in this case $V_{eff}(x) \sim V(x)$. Thus, hereafter we are going to use $r$ = -1 and $s$ = -3/2 ($\alpha = \gamma = -1/2, \beta = 0$) since in this case one has $V_{eff}(x) = V(x)$ and the Schrödinger equation is simply given by (29) and

$$\left(\hbar^2/2m_e\right)\left[d^2\phi/dx^2\right] = m_r(x)\left[E - V(x)\right]\phi = 0. \tag{37}$$

The mass dependent position function that we chose is

$$m_r(x) = 1 + \sum_{i=1}^{3} m_i \exp\left[-2\left(x - x_i\right)^2\right]. \tag{38}$$

Some eigenfuncions of (37)-(38) with $m_1 = m_2 = m_3 = 0.67$, $x_1$ = 4.5nm, $x_2$ = 10.5nm and $x_3$ = 16.5nm can be seen in Fig. 19. Differently of what has been observed, there exist modes in which the particle keeps confined only between the central barrier and a lateral barrier. The lowest order of those modes are $\psi_1$ (left quantum well) and $\psi_2$ (right quantum well).

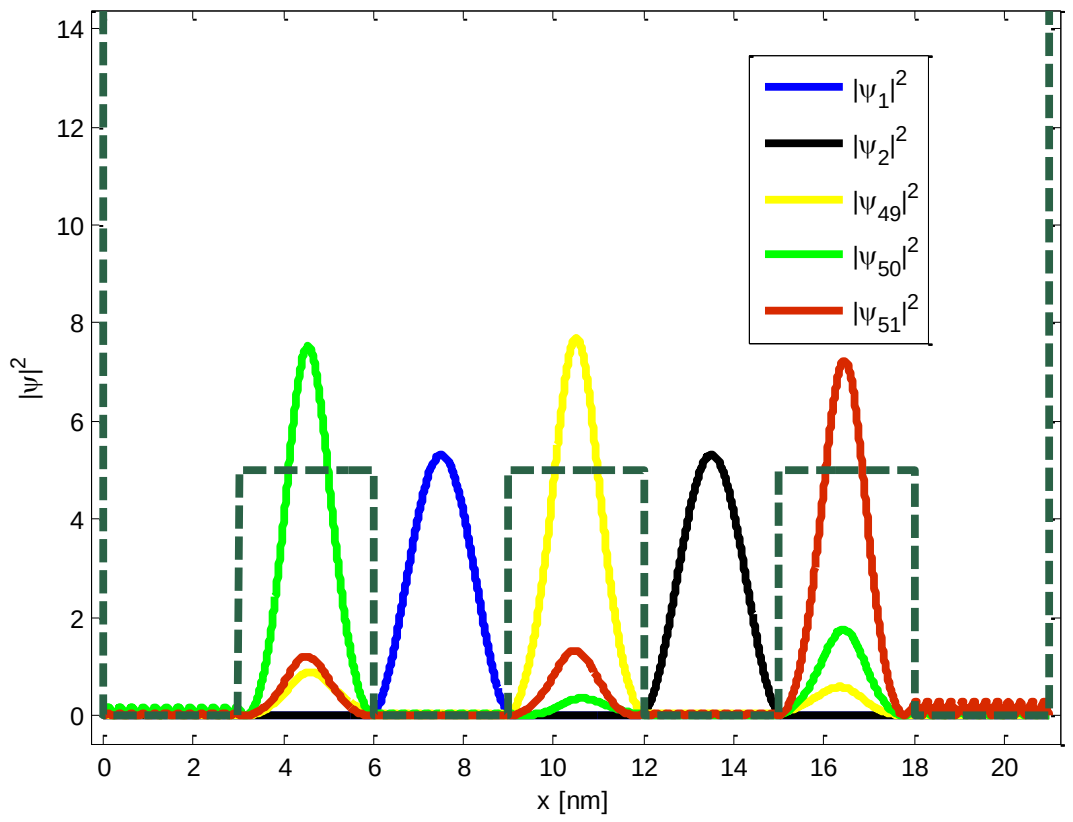

Fig. 19 – The squared modulus of the eigenfunctions $\psi_1$, $\psi_2$, $\psi_{49}$, $\psi_{50}$ and $\psi_{51}$. The parameters values are $V_1 = V_2$ = 5eV. The energy of the modes are $E_1$ = 0.0335eV, $E_2$ = 0.0335eV, $E_{49}$ = 5.0243eV, $E_{50}$ = 5.0285eV and $E_{51}$ = 5.0298.

Furthermore, the modes $\psi_{49}$, $\psi_{50}$ and $\psi_{51}$ represent the particle confined in the barrier regions. As Fig. 19 suggest, the allowed energies grow slowly when compared with the constant mass case. This can be seen in Fig. 20.

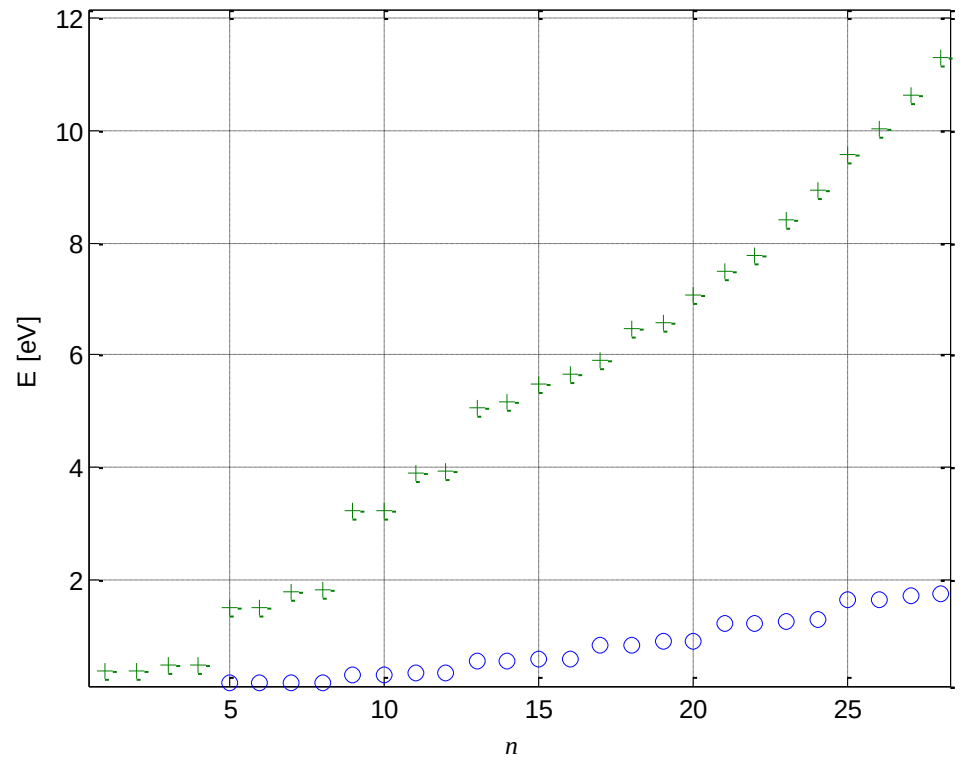

Fig. 20 – Energies of the first thirty modes. (+) DQW-TB with constant mass. (o) DQW-TB with dependent position mass given by (38).

## V. Conclusions

In this work we have been concerned with the solutions of the time independent 1D Schrödinger equation for the DQW-TB nanostructure. Two cases were analyzed: the scattering of a single-particle impinging in the DQW-TB, and the bound states and corresponding allowed energies for a DQW-TB inside of an infinite quantum well. Since in both cases there are a large number of parameters to be taken into account (shape, height and width of the three barriers, the distance between the barriers and the mass variation in the region considered) we analyzed some fixed structures letting free only few parameters, mainly the barriers' height. In the first case we used an analytical approach and we provided, for the first time in the best of our knowledge, a formula (Eq. (16)) for the transmission coefficient of a DQW-TB with twins lateral (rectangular) barriers. Regarding the DQW-TB inside of an infinite quantum well, we used the Numerov method to find eigenfunctions and the associated allowed energies. In this case there are four quantum wells formed by the walls of the infinite quantum well and the DQW-TB nanostructure, and the infinite quantum well by itself. For high values of allowed energies, the energies tend to follow the parabolic behaviour of the unperturbed infinite quantum well. On the other hand, for energies of the order and lower than the barriers height, that parabolic behaviour does not exist and modes mostly confined in the quantum wells (for energies lower than the barrier's height) and in the barriers' regions (for energies a little higher than the barriers' height) appear. For the mass constant case, the lowest energy mode is that one in which the particle is mostly confined in the regions between the central barrier and both lateral barriers. This quantum state suggests the particle can tunnel freely through the central barrier. This is the opposite of the lowest energy mode of the same structure with dependent position mass given by (38). There are two lowest energy modes and each one is mostly confined between the central barrier and a single lateral barrier, $\psi_1$ ($\psi_2$) between the central and the left (right) side barrier. It is also interesting to note that while the quantum wells confine the particle with energies lower than barriers' height, the barriers can also confine the particle when the energy is a little bit larger than the barriers' height.

This work was supported by the Brazilian agency CNPq via Grant no. 303514/2008-6. Also, this work was performed as part of the Brazilian National Institute of Science and Technology for Quantum Information.